\documentclass[a4paper,11pt]{article}
\pdfoutput=1 

\usepackage{jcappub} 

\usepackage[T1]{fontenc} 
\usepackage{multirow}
\usepackage{graphicx}

\begin{document}

\title{Circular orbits and accretion disk around AdS black holes surrounded by dark fluid with Chaplygin-like equation of state}                      

\author[a]{G. Mustafa,}
\author[b]{S. K. Maurya,}
\author[c]{A. Ditta,}
\author[d]{Saibal Ray,}
\author[e,f,g]{Farruh Atamurotov}

\affiliation[a]{Department of Physics, Zhejiang Normal University, Jinhua 321004, China}
\affiliation[b]{Department of Mathematics and Physical Science, College of Arts and Science, University of Nizwa, Nizwa, Sultanate of Oman}
\affiliation[c]{Department of mathematics, The Islamia University of Bahawalpur, Bahawalpur-63100, Pakistan}
\affiliation[d]{Center for Cosmology, Astrophysics and Space Science (CCASS), GLA University, Mathura 281406, Uttar Pradesh, India}
\affiliation[e]{New Uzbekistan University, Movarounnahr street 1, Tashkent 100000, Uzbekistan}
\affiliation[f]{Central Asian University, Milliy Bog' Street 264, Tashkent 111221, Uzbekistan}
\affiliation[g]{Institute of Theoretical Physics, National University of Uzbekistan, Tashkent 100174, Uzbekistan}

\emailAdd{gmustafa3828@gmail.com}
\emailAdd{sunil@unizwa.edu.om}
\emailAdd{adsmeerkhan@gmail.com}
\emailAdd{saibal.ray@gla.ac.in}
\emailAdd{atamurotov@yahoo.com}

\date{\today}

\abstract{In the present work we study the geodesic motion and accretion process of a test particle near an Anti-de Sitter (ADS) BH surrounded by a dark fluid with a Chaplygin-like equation. Within the defined paradigm, we investigate on the equatorial plane and examine circular geodesics along with their features related to stabilities, radiation energy flux, oscillations and orbits. The general form of the fluid accretion onto the AdS BH through dynamical analysis and mass expansion also has discussed in a depth. Additionally, a few more interesting topics, e.g. the effective potential, angular momentum, specific energy, radiation energy and epicyclic frequencies have also been examined thoroughly. All the attributes are physically acceptable within the observational signatures and ranges.}


\maketitle
\flushbottom

\section{Introduction} ~\label{Intro} 
As a theory of gravitation Einstein's General Relativity (GR) provides an in-depth and wide infrastructural facilities to understand all kinds of phenomena involved in the field of astrophysics and cosmology. Remarkably, Schwarzschild~\cite{Schwarzschild1916} first found out the solution to GR in connection to stellar model which can be treated as a toy model devoid of most of the real situations and physical aspects. Nevertheless, this historical solution provided an interesting feature of GR that this theory can conceive the idea of black hole (BH) in the spherically symmetric static metric. Here one may face an awkward situation for the condition either $R=2M$ or $r \rightarrow 0$, where $r$, $R$ and $M$, respectively are the radial coordinate, the radius of the stellar system and mass of the star. This odd and unexpected feature in the Schwarzschild~\cite{Schwarzschild1916} is the well-known {\it singularity} of a BH. At this juncture it may be a funny information that Einstein did not believe even in the existence of black holes where he refutes it by commenting that ``Schwarzschild singularities" do not exist in physical reality~\cite{Einstein1939,Bernstein2007}.

However, the historical timeline how the ``Schwarzschild singularities" gradually have taken shape over the time are as follows: (i) in 1939, Oppenheimer along with Volkoff worked on the collapsing star and possibility of the formation of a BH~\cite{OV1939}, (ii) again in the same year Oppenheimer and Snyder proposed a BH scenario in their paper `On Continued Gravitational Contraction'~\cite{OS1939}, (iii) in 1958, Finkelstein pointed out the Schwarzschild surface as an event horizon~\cite{Finkelstein1958}, (iv) in 1960, an extension of the Schwarzschild model was proposed by Kruskal~\cite{Kruskal1960}, (v) in 1963, Kerr obtained an exact solution for a rotating BH~\cite{Kerr1963}, (vi) in 1965, Newman~\cite{Newman1965} investigated on the axisymmetric solution for a BH with rotation and charge, (vii) in 1965, Penrose~\cite{Penrose1965} found generic solutions to Einstein's GR which are involved in singularities, (viii) in 1967-1977, through a series of research papers the no-hair theorem emerged out which attributes that a stationary BH solution can completely be described by the three physical parameters, viz. mass, angular momentum and electric charge~\cite{Israel1967,Carter1971,Robinson1975,Carter1977}, (vii)  (vii) in the period 1973-1975, a thorough investigations by Bardeen, Carter and Hawking~\cite{BCH1973} and Bekenstein~\cite{Bekenstein1973,Bekenstein1974,Bekenstein1975} formulated the BH thermodynamics, (ix) in 1974, Hawking~\cite{Hawking1974} raised an novel question whether "Black hole explosions?" takes place which eventually gave birth to the so-called {\it Hawking radiation} and (x) innumerable theoretical studies on BH by hundreds of scientists, recently studied a few of them are as follows~\cite{Myung2009,Liang2012,Habib2012,Ghosh2012,Rahaman2013,Radinschi2013,Bhattacharyya2022a,Bhattacharyya2022b,Tello-Ortiz12023}. Most recently, some new black hole solutions have been calculated in the background of different aspect including gravitational decoupling approach and quantum induced ~\cite{rjim1,rjim2,rjim3,rjim4,rjim5,rjim6,rjim7,rjim8,rjim9,rjim10,rjim11,rjim12,rjim13,rjim14,rjim15}

On the other hand, based on the observational evidence, in 1975, a galactic X-ray source named as Cygnus X-1, has been recognized as the first black hole~\cite{SYD1975,Rolston1997}. However, very recently on 11 February 2016, Abbott et al.~\cite{Abbott2016} announced the first direct detection of gravitational waves which represents the observation of a BH merger. Again on 10 April 2019, the first direct image of a BH and its vicinity was published following observations made by the Event Horizon Telescope (EHT) in 2017 of the supermassive black hole in Messier 87's galactic centre~\cite{Bouman2016,EHT2019}. Till date there have been collective evidences on the real existence of BH~\cite{Kareem2023}. 

In connection to major properties and structures, whether it is of primordial or stellar or galactic center-based BH, there are a lot of works available in the literature, viz. (i) the Hawking radiation which has already been mentioned~\cite{Hawking1974} is a unique combination of GR and quantum mechanical effect, (ii) the Lense–Thirring effect which becomes indicative due to the angular momentum related frame dragging by the gravitomagnetic field~\cite{Reynolds2019}, (iii) an event horizon which acts as an one-way boundary membrane in spacetime and allows matter as well as light-quanta to pass inside~\cite{Davies1992,Fleisch2013,Wheeler2007}, (iv) the innermost stable circular orbit is a special feature of GR for which any infinitesimal inward perturbations to a circular orbit will lead to spiraling into the black hole~\cite{MTW1973} and (v) an accretion disk which is a structural formation around a BH by diffused material in order to maintain angular momentum~\cite{Weizsacker1948}. The impact of a cosmological constant on the test-particle motion in BH backgrounds was examined by Stuchlík \cite{rr55}. Stuchlík computed the circular orbits and the characteristics of purely radial trajectories around the charged black hole while accounting for the cosmological constant in the same study. In a different study, Schwarzschild-de Sitter and Schwarzschild-anti-de Sitter spacetime properties were examined by Stuchlík and his collaborator Hledík \cite{rr56}. They made use of the embedding diagrams, photon escape cones, and effective potential for the motion of test particles and photons. They demonstrate that in the Schwarzschild–de Sitter spacetimes, coincidence does not hold. In a recent review article, Stuchlík et al., \cite{rr58} examined the effects of cosmic repulsion and external magnetic fields on jets connected to rotating BHs and accretion disks rotating around them against the background of the cosmological constant. 

In the present work related to BH we are interested to focus on the two specific physical properties of it, viz. circular orbits and accretion disks. Regarding the circular orbits, the basic idea is that in GR there does exist an innermost stable circular orbit contrary to the Newtonian gravity where a test particle can orbit steadily in reference of a central massive object. Depending on the situations, i.e. energy of the test particle and spin of the BH, a circular orbit may lead to spiraling effect~\cite{MTW1973}. As far as inward spiraling of a particle is concerned one can have a proximity to an accretion disc which is a resultant of accumulative particles due to tremendous gravitational attraction acting on the gaseous matter. 

Therefore, based on the above mentioned literature and physical status of some of the astrophysical aspects, our attempt in the present article is to carry out investigation on some features of BH, especially circular orbits and accretion disks. 

To reach at this goal we have presented first an AdS BH solution by considering the spherically symmetric spacetime in Section~\ref{sec2}. In the next Section~\ref{sec3} we have provided geodesic motion regime under which circular motions (\ref{Circular}), radiation energy flux (\ref{Radiation}) and oscillations (\ref{Oscillations}) have been discussed. For the accretion disks related discussion we have chosen Section~\ref{sec4} where we have presented dynamic parameters~(\ref{Dynamic}) and mass expansion~(\ref{Mass}). We provide a brief overview of the circular equatorial geodesics~\ref{sec5} under which we specifically presented epicyclic frequencies~(\ref{Epicyclic}). Finally, Section~\ref{sec6} is devoted to for discussion and concluding remarks.

\section{AdS black hole surrounded by dark fluid with Chaplygin-like equation of state}~\label{sec2} 

To investigate the connection between the BH spacetime and the field when the BH is static spherically symmetric and its atmosphere is made up of a field with an explicit Lagrangian, it is practical to solve the gravitational field equations and the equation of motion of the relevant field simultaneously. When we are unable to understand the nature of the matter around the BH solutions due the influence of  quintessence dark energy and Chaplygin gas, then it is crucial to think about how to investigate the relationship between matter and spacetime curvature using just the fluid matter's equation of state. For the current analysis, We consider the following form for the metric to describe our static spherically symmetric spacetime.
\begin{equation}\label{eq1}
ds^{2}=-f(r)dt^{2}+\frac{1}{f(r)}dr^{2}+r^{2}(d\theta^{2}+ \sin^{2}\theta d\phi^{2}).
\end{equation}

In order to matter source, we start with the perfect fluid, which is given as
\begin{equation}\label{eq2}
T_{ab}=pg_{ab}+ (\rho+p) u_a u_b, 
\end{equation}
where $p$ and $\rho$ mention the isotropic pressure and energy density respectively. 

As pioneer in the working field, Semiz ~\cite{jim1} has investigated the equation of state (EOS) $p=\omega \rho$ ($\omega$ is a constant) for ideal fluid sources (dust, phantom energy, radiation, or dark energy) by using Einstein's equations as a basis. Further, Kiselev~\cite{jim2} calculated a novel BH spacetime by considering the ambient quintessence matter as an anisotropic fluid, which quickly turned into an extraordinarily popular toy model. For the Chaplygin dark fluid (CDF), from a field theoretical perspective, even now its background producing processes remain unclear~\cite{Hulke2020,Ray2023}, while there are a few plausible options in the framework of string theory~\cite{Nozari2011,Benaoum2012}. The CDF is typically represented by adding a self-interacting potential and a scalar field $\varphi$ with the Lagrangian $L_{\varphi}=-\frac{1}{2}\partial_{a}\varphi\partial^{a}\varphi-U(\varphi)$~\cite{jim3,jim4}. In this study, we assume that the CDF is anisotropic and that a covariant form of its stress-energy tensor may be expressed as~\cite{jim5}
\begin{equation}\label{eq3}
T_{ab}= \rho u_a u_b +p_r k_a k_b +p_t \Pi_{ab}, 
\end{equation}
where $p_r$ and $p_t$ are defining the radial and the tangential pressure components, $u_{a}$ defines the four-velocity of fluid and $k_{a}$ represents a unit space-like vector, which is orthogonal to $u_{a}$, with $u_{a}$ and $k_{a}$ with satisfying \mbox{ $u_{a} u^{\mathrm{a}}=-1$}, \mbox{ $k_{a}k^\mathrm{a}=1$} and $u^{\mathrm{a}}k_{a}=0$. $\Pi_{ab} = g_{ab}+u_{a}u_{b}-k_{a}k_{b}$ is a tensor projection onto two orthogonal surfaces $u^\mathrm{a}$ and $k^\mathrm{a}$. When one operates within the fluid's comoving frame, they can achieve that $u_{a} = (-\sqrt{f},0,0,0)$ and $k_{a} = (0,1/\sqrt{g},0,0)$. 

\begin{figure}[!htp]
\centering \epsfig{file=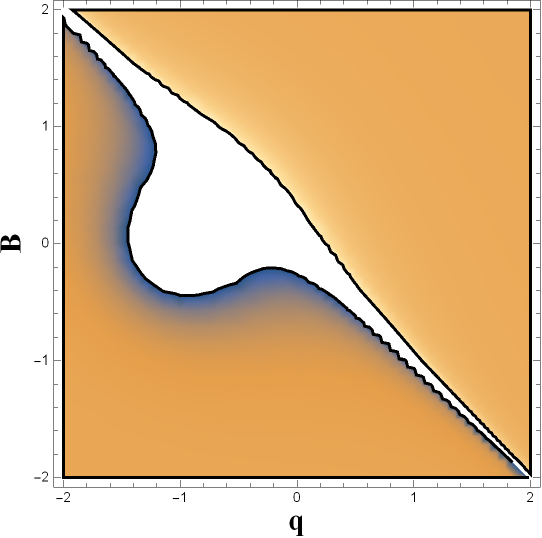, width=.45\linewidth,
height=2.3in}\caption{\label{F1} An illustration of the BH mass varying with $q$ and $B$. We set $\Lambda$ = -1. A physical BH mass needs to match the extremal BH mass at least, for the specific values of $B$, $q$, and $\Lambda$.}
\end{figure}

Finally, the stress-energy tensor through Eq.~(\ref{eq3}) can be provided as
\begin{equation}
T_{a}{}^{b}=-(\rho+p_t)\delta_{a}{}^0\delta^{b}{}_0 +p_t \delta_{a}{}^{b} +(p_r-p_t)\delta_{a}{}^1\delta^{b}{}_1. \label{eq4}
\end{equation}
The anisotropic factor is defined as the difference between radial and tangential pressures $p_r-p_t$. The stress-energy tensor simplifies to the conventional isotropic at $p_r=p_t$. Even if a cosmological fluid has anisotropy in the gravitational field generated by a BH, its equation of state should appear as $p=p(\rho)$ at cosmological scale, allowing one to constrain the tangential pressure $p_t$ by taking an isotropic average over the angles and and some extra conditions $\langle {T}_i{}^j\rangle=p(\rho)\delta_i{}^j$, one can get the following relation
\begin{equation}
p(\rho)=p_t+\frac{1}{3}(p_r-p_t),\label{eq4a}
\end{equation}
where the relation $\frac{1}{3}=\langle \delta_i{}^1\delta^j{}_1\rangle$ has been considered. 
The EOS $p =\omega\rho$ ($-1<\omega<-1/3$) for quintessence matter, from the Eq.~(\ref{eq4a}), one can use $p_t=\frac{1}{2}(1+3\omega)\rho$, which is compatible with the radial pressure $p_r=-\rho$~\cite{jim2}. The non-linear EOS for the CDF in our scenario is $p=-\frac{B}{\rho}$, where $B$ is a positive constant. The tangential pressure produces $p_t=\frac{1}{2}\rho-\frac{3B}{2\rho}$ for $p_r=-\rho$. As a result, the CDF's stress-energy tensor can be written as
\begin{eqnarray}
&-\rho={{T}_t}^t={{T}_r}^r, \label{Ttr}~\\
&\frac{1}{2}\rho-\frac{3B}{2\rho}={{T}_{\theta}}^{\theta}={{T}_{\phi}}^{\phi}.~\label{Tangular}
\end{eqnarray}
At cosmological scale, the anisotropy of the CDF vanishes and the EOS yields $p=-B/\rho$. Moreover, as an important condition, i.e., ${{T}_t}^t={{T}_r}^r$ is required, the relationship between the metric components $g(r)=f(r)$ can be accomplished through a suitable time rescaling without sacrificing generality. The Einstein tensor components are calculated as
\begin{eqnarray}
&{{G}_r}^r=\frac{1}{r^2}(f+rf'-1)={{G}_t}^t,~\label{Gtr}\\
&\frac{1}{2r}(2f'+rf'')={{G}_{\theta}}^{\theta}={{G}_{\phi}}^{\phi}. \label{Gthetafai}
\end{eqnarray}
Now, by using the Eqs.~(\ref{Ttr})-(\ref{Tangular}) and Eqs.~(\ref{Gtr})--(\ref{Gthetafai}), one can get the following relation
\begin{eqnarray}
&-\rho=\frac{1}{r^2}(f+rf'-1)+\Lambda,~\label{graviequationsrt}\\
&\frac{1}{2}\rho-\frac{3B}{2\rho}=\frac{1}{2r}(2f'+rf'')+\Lambda.~\label{graviequationsthetafai}
\end{eqnarray}
We now examine the existence of the cosmological constant. The two above differential equations allow us to calculate the two unknown functions, i.e., $f(r)$ and $\rho(r)$, analytically. Now, one can simply the energy density of CDF by solving the above set of differential equations (\ref{graviequationsrt})--(\ref{graviequationsthetafai}) as:
\begin{equation}
\rho(r)=\sqrt{B+\frac{q^2}{r^6}}, \label{CDFenergydensity}
\end{equation}
where a normalization factor $q>0$ denotes the CDF's intensity. Additionally, the conservation law for the stress-energy tensor $\nabla_{b}T^{ab}=0$ directly yields Eq.~(\ref{CDFenergydensity}). It is evident that the CDF energy density can be approximated by $r^6\ll q^2/B$ as
\begin{equation}
\rho(r)\approx\frac{q}{r^3}, \label{CDFenergydensitysmallr}
\end{equation}
demonstrating that the CDF behaves as though its energy density were dependent on $r^{-3}$ for matter content. Now, by using Eq.~(\ref{CDFenergydensity}) into Eq.~(\ref{graviequationsrt}), we have the final analytical solution for $f(r)$ ~\cite{jim6} as
\begin{equation}
f(r)=1-\frac{2M}{r}+\frac{q}{3r}{\rm ArcSinh}\frac{q}{\sqrt{B}r^3}-\frac{r^2}{3}\sqrt{B+\frac{q^2}{r^6}}-\frac{ r^2}{3}\Lambda, \label{frsolution}
\end{equation}
where $M$ is the BH mass, however in the current case, the BH mass is regarded as a point mass BH.  Given $r\rightarrow \infty$, $f(r)$ can be adopted asymptotically
\begin{eqnarray}\label{3}
f(r)&\rightarrow&1-\frac{r^2}{3}(\sqrt{B}+\Lambda).
\end{eqnarray}
The solution of the asymptotic behavior is examined by both the CDF parameter $B$ and cosmological parameter $\Lambda$. In this work, we apprehension the
AdS BH, so we constrain $\Lambda<-\sqrt{B}$. Fig. (\ref{F1}) shows how the metric function $f(r)$ depends on the parameters $B$ and $q$.

To study the horizon structure, one can take $f(r_h)=0$, and find the event horizon radius $r_h$, with the mass of the BH can be expressed as
\begin{eqnarray}\label{4}
M&=&1-\frac{r_h}{2}-\frac{r^3_h}{6}\Lambda-\frac{r^3_h}{6}\sqrt{B+\frac{q^2}{r^6_h}}+\frac{q}{6}ArcSinh\frac{q}{\sqrt{B}r^3_h}.
\end{eqnarray}

\begin{figure}
\centering \epsfig{file=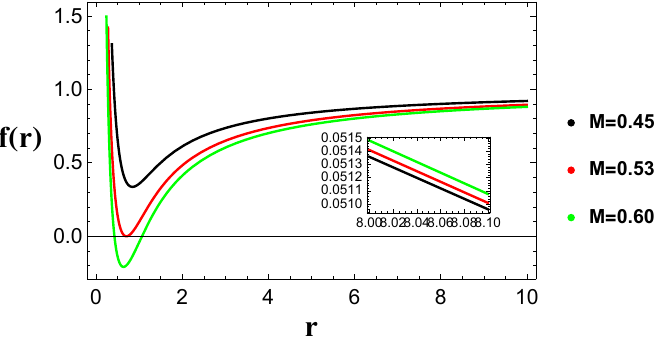, width=.47\linewidth,
height=2.05in}~~~~~~\epsfig{file=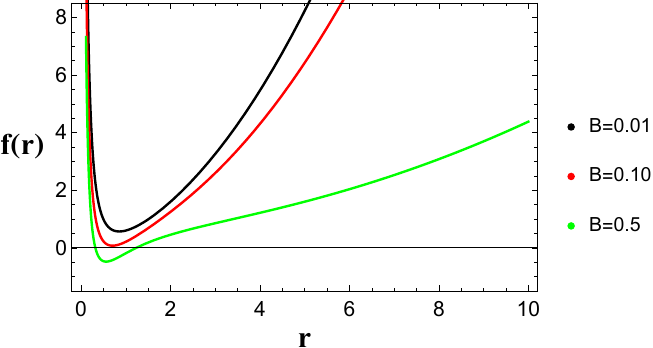, width=.47\linewidth,
height=2.05in}
\centering \epsfig{file=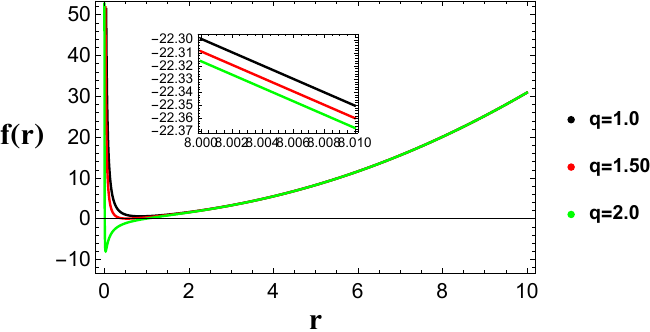, width=.47\linewidth,
height=2.02in}\caption{\label{F2} An illustration of horizons of the AdS BH $(f(r))$ versus $r$ for different values of $M$, $B$ and $q$.}
\end{figure}

The effect of parameters on the lapse $f(r)$ is shown in Fig. (\ref{F2}). Some important insights regarding horizon structure are listed below:
\begin{itemize}

  \item In the upper plot by the left position has three solution curves black,red and green, in which black curve has no horizon for $M=0.45$. 
  Red curve has an event horizon for $M=0.53$ and green curve has two horizon for $M=0.60$ at the specific values of other parameters $B=1$, $q=1$ and $\Lambda=-1$.
  
  \item The upper right plot has three solution curves black,red and green, in which black curve has no horizon for $B=0.01$.
  Red curve has an event horizon for $B=0.10$ and green curve has two horizon for $B=0.50$ at the specific values of other parameters $M=1$, $q=1$ and $\Lambda=-1$.
  
  \item The bottom plot has three solution curves black,red and green, in which black curve has no horizon for $q=1.0$.
  Red curve has an event horizon for $q=1.50$ and green curve has two horizon for $q=2.0$ at the specific values of other parameters $M=1$, $B=1$ and $\Lambda=-1$.
  
\end{itemize}
We have noted the interesting behavior of AdS BH, all the solution curves inclines towards the event horizon as all the parameters are increased in value of $M$, $B$ and $q$.

\section{Geodesic motion regime}~\label{sec3} 

In this Section, we compute the overall findings for particle geodesic motion by using the underlying symmetric and static space-time. For the analysis, we make a choice of two Killing vectors, say $\varepsilon_t=\partial_t$ and $\varepsilon_\phi=\partial_\phi$ under the effect of two constants of motion like $E$ (conserved energy) and other one is $L$ (angular momentum) with the following formulas:
\begin{eqnarray}\nonumber
E&=&-g_{ab}\varepsilon^a _t u^b\equiv-u_t,\\
\label{9}
L&=&g_{ab}\varepsilon^a _\phi u^b\equiv u_\phi,
\end{eqnarray}
with $u^a = \frac{dx^a}{d\tau}=(u^t,u^r,u^\theta,u^\phi)$ are the four-velocity vectors for the moving particles, which fulfil the normalization condition, i.e., $ u^a u_a = -1$, one can get the following relation:
\begin{equation}
g_{rr}(u^r)^2+g^{tt}(u_t)^2=-\left(1-g_{\theta\theta}(u^\theta)^2-g^{\phi\phi}(u_\phi)^2\right).\label{10}
\end{equation}

Along the equatorial plane, i.e., ($\theta=\frac{\pi}{2}$), from Eqs. (\ref{9}) and (\ref{10}), we have the following relations
\begin{eqnarray}\label{11}
u^t&=&\frac{E}{f(r)},\\\nonumber
u^\theta&=&0,\\\nonumber
u^\phi&=&\frac{L}{r^2},\\\nonumber
u^r&=&\sqrt{f(r)\left(-1+\frac{E^2}{f(r)}-\frac{L^2}{r^2}\right)}.
\end{eqnarray}
From Eq. (\ref{11}), the conserved energy equation for the particles motion can be obtained with an effective potential $V_{eff}$ as
\begin{equation}
E^2=(u^r)^2+V_{eff}.\label{12}
\end{equation}
\begin{equation}
V_{eff}=f(r)\left[1+\frac{L^2}{r^2}\right].\label{13}
\end{equation}

\begin{figure}
\centering \epsfig{file=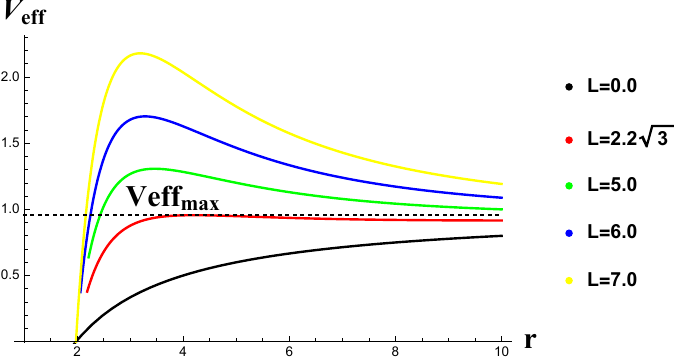, width=.45\linewidth,
height=2.05in}~~~~~~~~\epsfig{file=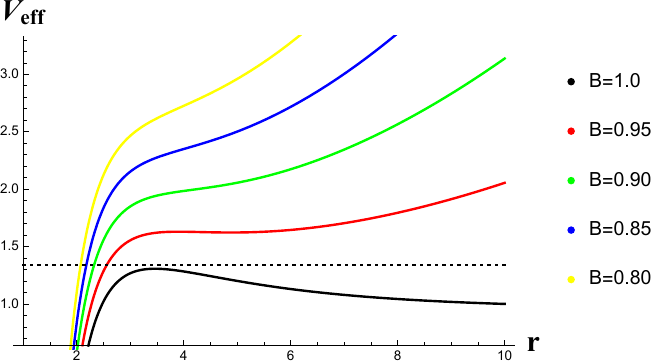, width=.45\linewidth,
height=2.05in} \\
\epsfig{file=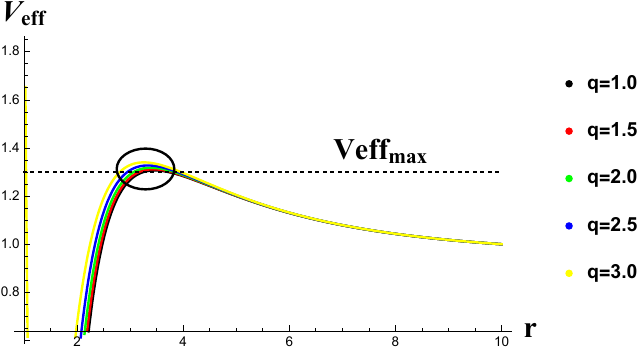, width=.45\linewidth,
height=2.02in}\caption{\label{F3} An illustration of $V_{eff}$ of the AdS BH versus $r$ for different values of $B$ and $q$.}
\end{figure}

The effective potential analysis of AdS BH plays a crucial role in geodesic motion, which clearly depends on parameter $f(r)$ and the angular momentum $L$. The local extremum of the effective potential can be used to identify the position of the circular orbits. The following are the salient features of Fig. \ref{F3}:

\begin{itemize}
  \item In the upper plot by the left position, the black curve has no extremum for $L=0$. The first extremum is detected in red curve at $V_{eff}=0.99$ for $L=2.2\sqrt{3}$. In this plot $V_{eff}$ is increased for increasing $L$. In this plot, the values of other parameters are $M=1$, $\beta=1$, $q=1$ and $\Lambda=-1$.       
  \item In the left right plot, the black curve has an extremum for $B=1.0$. The $V_{eff}$ is increased for decreasing $B$. In this plot, the values of other parameters are $M=1$, $L=5$, $q=1$ and $\Lambda=-1$.  
  \item In the bottom plot, the black curve has an extremum for $q=1.0$. The solution curves of $V_{eff}$ are very small increased for increasing $q$.
  In this plot, the values of other parameters are $M=1$, $\beta=1$, $L=5$ and $\Lambda=-1$.
\end{itemize}
We investigated the position of the orbits, which is represented by the point of extremum around the AdS BH at a distance of $r \approx 2$, where the potential increases for increasing the angular momentum $L$ and parameter $q$ decreases for reducing the parameter $B$.

\subsection{Circular motions}~\label{Circular} 
An equatorial plane can be used to investigate a particles circular motion. In this case, the radial component of circular motion $r$ must be constant, i.e. $u^r=\dot{u}^r=0$. We get $V_eff=E^2$ and $\frac{d}{dr}V_eff=0$ from Eq. (\ref{12}).  These relationships allow us to write the particular energy $E$, specific angular momentum $L$, angular velocity $\Omega_\phi$, and angular momentum $L$ as:
\begin{eqnarray}\label{14}
E^2&=&\frac{2f^{2}(r)}{-r f'(r)+2f(r)},\\
\label{15}
L^2&=&\frac{r^2 f'(r)}{2f(r)-rf^{'}(r)},\\
\label{16}
\Omega_{\phi}&=&\frac{d\phi}{dt}\equiv\frac{u^\phi}{u^t}\Rightarrow \Omega^2_{\phi}=\frac{f'(r)}{2r},\\
\label{17}
l^2&=&\frac{L^2}{E^2}=\frac{r^3f'(r)}{2f^2(r)}.
\end{eqnarray}
To achieve the specific energy and angular momentum, we have the following condition
\begin{equation}
r\left(2f(r)-rf'(r)\right)>0.\label{18}
\end{equation}
One can acquire the limiting area of a circular orbit by solving the above condition. Additionally, two relations $E^2<1$ and $E^2=1$ are taken for the bound and marginally bound orbits respectively. Marginally bound orbits are easily discovered by using Eq. (\ref{14}). Thus 
\begin{equation}
r\left(2f(r)[f(r)-1]+rf'(r)\right)=0.\label{19}
\end{equation}
By using Eqs. (\ref{14}) and (\ref{15}), the momentum and energy will diverge at the radius $r$. The divergence relation between Eqs. (\ref{14}) and (\ref{15}) is provided as
\begin{equation}
r\left(-r^2f'(r)+2f(r)\right)=0.\label{20}
\end{equation}
The photon sphere is characterized by this relation.

\begin{figure}
\centering \epsfig{file=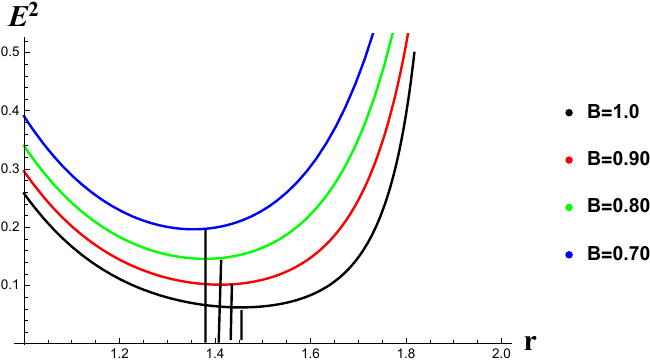, width=.45\linewidth,
height=2.05in}~~~~~~~~\epsfig{file=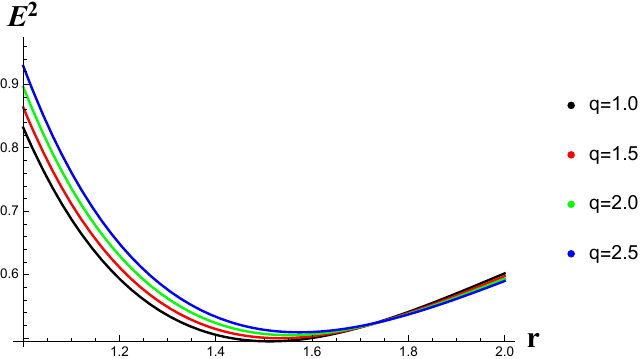, width=.45\linewidth,
height=2.02in}
\caption{\label{F4} The illustration of the specific energy $E$ of the AdS BH versus $r$ for different values of $B$ and $q$.}
\end{figure}

\begin{figure}
\centering \epsfig{file=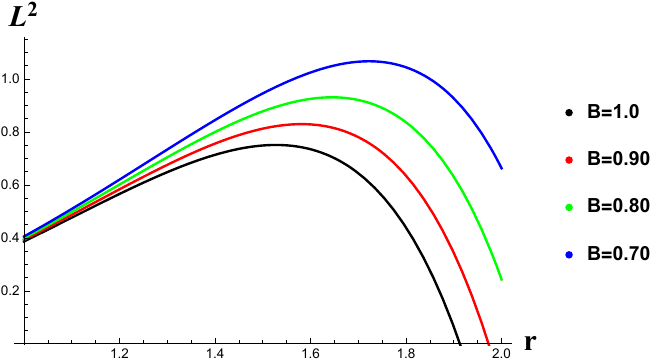, width=.45\linewidth,
height=2.05in}~~~~~~~~\epsfig{file=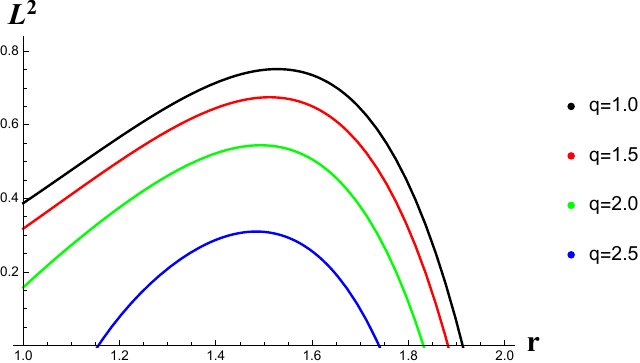, width=.45\linewidth,
height=2.02in}\caption{\label{F5} The illustration of the angular momentum $L$ of the AdS BH versus $r$ for different values of $B$ and $q$.}
\end{figure}
The significant points of the specific energy, which is shown in Fig. \ref{F4} are as follows:
\begin{itemize}
  \item The specific energy is decreased gradually by increasing the parameter $B$. The curves are shifted inwards to the singularity.
 \item The specific energy is decreased gradually by decreasing the parameter $q$ and vice versa. The curves are shifted inwards to the singularity until it hits the singularity at $r=1.5$ and go up to the maximum radius.
  \end{itemize} 
The angular momentum was described in Fig. \ref{F5}, and its salient features are as follows:
\begin{itemize}
  \item The angular momentum is decreased gradually by increasing the parameter $B$. The curves are shifted onward to the singularity. However, the black and red curves are quickly fall onto the singularity and decreasing the radius.
\item The angular momentum is decreased gradually by increasing the parameter $q$ and vice versa. All the curves quickly fall onto the singularity and decreasing the radius and vice versa.
\end{itemize}

\subsection{Radiation energy flux}~\label{Radiation}
The most explosive astrophysical events can be caused by radiation that is created when falling elements that are falling from rest at infinity accrete onto the black hole. This process releases gravitational energy from the falling elements. The radiation energy flux over the accretion disk is caused by the radiant energy associated with the specific energy $E$, angular momentum $L$ and angular velocity $\Omega_\phi$ that were examined by Kato et al.~\cite{Kato2008:10} and can be provided as 
\begin{equation}
K=-\frac{\dot{M}\frac{d\Omega_\phi}{dr}}{4\pi \sqrt{-g}(-L\Omega_\phi+E)^2}\int(-L\Omega_\phi+E)\frac{d}{dr}Ldr,\label{22}
\end{equation}
where $\dot{M}$ is an accretion rate, $K$ is a radiation flux and $g$ is another important parameter, which is defined as
\begin{equation}
g=det(g_{ab})=-r^4 \times \sin^2 \theta.\label{23}
\end{equation}

On the equatorial plane, one can obtain the following relation
\begin{eqnarray}\label{24}
K(r)=&&-\frac{\dot{M}}{4\pi r^4}\frac{r}{\sqrt{2f'(r)}}
\\\nonumber&&\times \frac{W X}{Y^2}
\int^{r}_{mb}Z(r)dr,
\end{eqnarray}
where $W =\left(-r f'(r)+2f(r)\right)$, $X =\left(-f'(r)+r f''(r)\right)$, $Y =\left(r f'(r)+2f(r)\right)$ and
\begin{eqnarray}\label{25}
Z(r)=&&\sqrt{\frac{r}{2f'(r)}}\frac{Y[-f''(r)r f(r)-3f'(r)f(r)+2r f'^2(r)]}{W^2}.
\end{eqnarray}

Now, by considering the relation $K(r)=\sigma T^4(r)$, between the radiation and temperature. Here, $\sigma$ is Stefan's constant. The detailed scenario of this Stefan's constant was discussed by~\cite{Torres2002:51} and further studied by~\cite{Babichev2005:52}.

\begin{figure}
\centering \epsfig{file=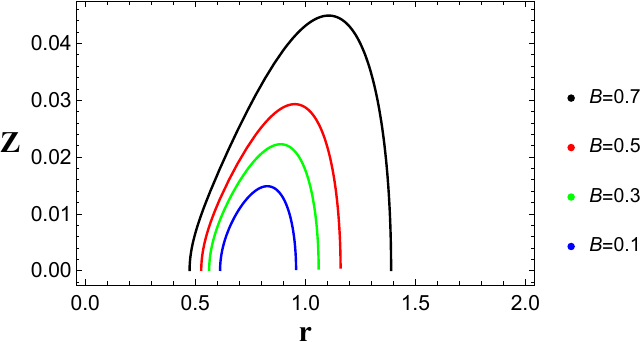, width=.45\linewidth,
height=2.05in}~~~~~~~~\epsfig{file=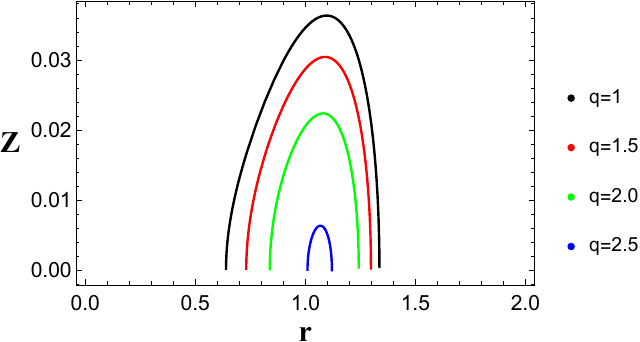, width=.45\linewidth,
height=2.02in}\caption{\label{F6} An illustration of $Z$ of the AdS BH versus $r$ for different values of $B$ and $q$.}
\end{figure}

\begin{figure}
\centering \epsfig{file=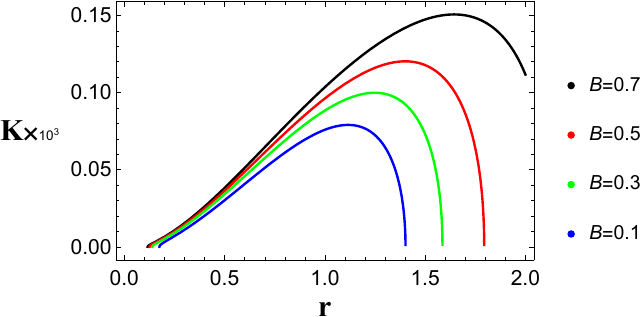, width=.45\linewidth,
height=2.05in}~~~~~~~~\epsfig{file=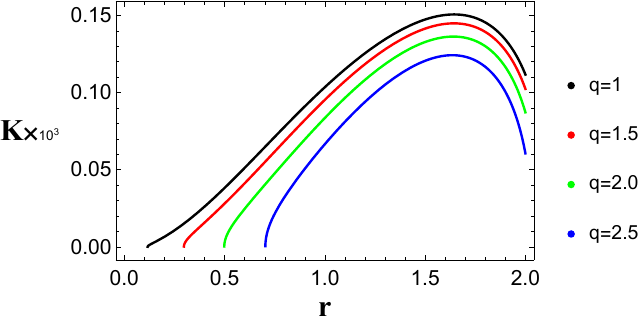, width=.45\linewidth,
height=2.05in}\\ \hspace{4cm}
\centering \epsfig{file=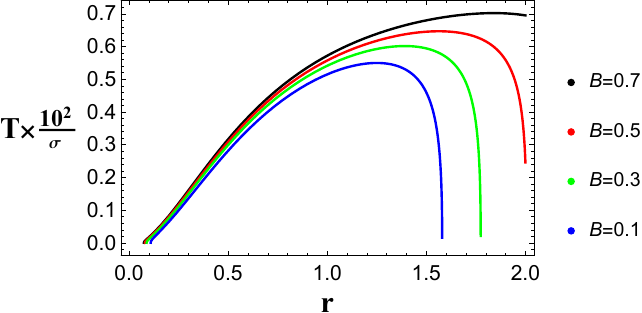, width=.45\linewidth,
height=2.02in}~~~~~~~~\epsfig{file=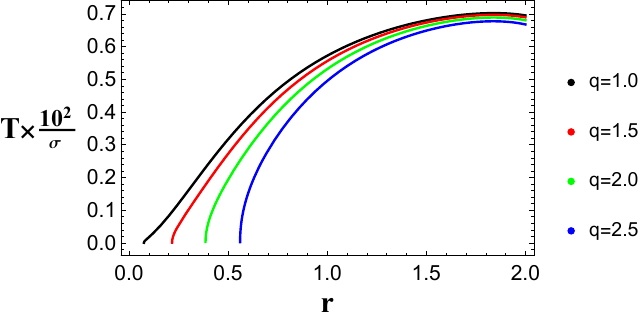, width=.45\linewidth,
height=2.02in}\caption{\label{F7} An illustration of the radiation energy $K(r)$ and the temperature of the AdS BH versus $r$ for different values of $B$ and $q$.}
\end{figure}

We described the radiation energy flux in Figs. (\ref{F6}) and (\ref{F7}).

\begin{itemize}
  \item The black curve has a maximum radiation energy flux for $B=0.70$ and $q=1.0$ and it is decreasing for decreasing values of $B$ and increasing $q$. The radius is decreased as the radiation flux increased by the parameters.
  
  \item The same behavior is seen in the picture of temperature and the radiation parameter $Z$ as seen in the radiation energy flux.
\end{itemize}

\begin{figure}
\centering \epsfig{file=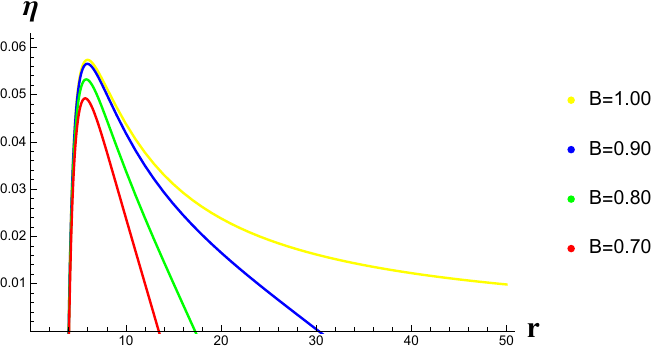, width=.45\linewidth,
height=2.05in}~~~~~~~~\epsfig{file=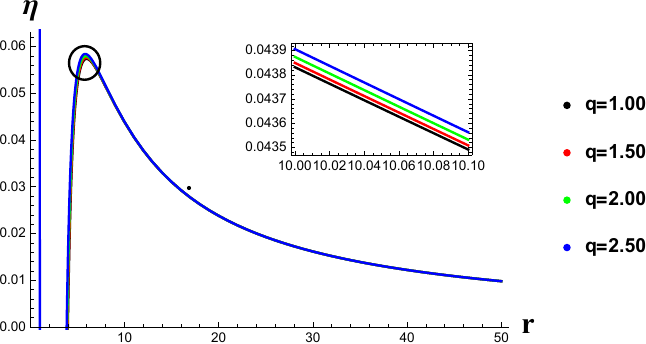, width=.45\linewidth,
height=2.02in}\caption{\label{F8} An illustration of the efficiency of the AdS BH versus $r$ for different values of $B$ and $q$.}
\end{figure}

An additional essential evaluation of the accretion disk is the efficiency of the accreting particle. Consequently, to achieve the highest accreting efficiency, we use the relation $\eta=1-E$. We described the accreting efficiency in Fig. \ref{F8} and has the following key insights:

\begin{itemize}
  \item The efficiency is increased for increasing both the parameters $B$ and $q$. 
  
  \item Increasing efficiency increased the radius and the curves are go up to the maximum radius.
  
  \item We see the increasing efficiency and vice versa in left plot for taking specific value of $q=1$ and variation in $B$.
  
  \item We see the increasing efficiency and vice versa in right plot for taking specific value of $B=1$ and variation in $q$.
  
\end{itemize}

\subsection{Oscillations}~\label{Oscillations}

In general, three frequencies are needed to discuss the particle motion in an accretion disk. Harmonic motion within the vertical frequency,  $\Omega_\theta$, harmonic motion within the radial frequency, $\Omega_r$, and harmonic motion within the orbital frequency $\Omega_\phi$. Here, the radial and vertical motions over the circular equatorial plane can be calculated. In this regard, we examine the following relations $\frac{1}{2}\left(\frac{dr}{dt}\right)^2=V^{(r)}_ {eff}$
and $\frac{1}{2}\left(\frac{d\theta}{dt}\right)^2=V^{(\theta)}_ {eff}$ respectively. By using the Eq. (\ref{9}), one can get radial motion $u^\theta=0$ and vertical motion, $u^r=0$. Further, using $u^r=\frac{dr}{d\tau}=\frac{dr}{st}u^t$ and $u^\theta=\frac{d\theta}{d\tau}=\frac{d\theta}{st}u^t$, we have
\begin{eqnarray}\label{26}
\frac{1}{2}\left(\frac{dr}{dt}\right)^2&=&-\frac{1}{2}\frac{f^3 (r)}{E^2}\left[1+\frac{E^2}{f(r)}+\frac{L^2}{r^2\sin^2 \theta}\right]=V^{(r)}_ {eff}.\\\nonumber
\frac{1}{2}\left(\frac{d\theta}{dt}\right)^2&=&-\frac{1}{2}\frac{f^2(r)}{r^2E^2}\left[1+\frac{E^2}{f(r)}+\frac{L^2}{r^2\sin^2 \theta}\right]=V^{(\theta)}_ {eff},
\end{eqnarray}
where $\Omega^2 (\theta)=-\frac{d^2}{d\theta^2}V^{\theta}_ {eff}$. Then from Eq. (\ref{26}), we obtain
\begin{equation}
\Omega^{2}_{\theta}=\frac{f^2 (r)L^2}{r^4 E^2}.\label{30}
\end{equation}
and
\begin{eqnarray}\label{31}
\Omega^{2}_{r}=&&\frac{1}{2r^4 E^2}[(L^2+r^2)r^2f^2(r)-f(r)r^4E^2]f''(r)\\\nonumber
&&-2r\left[[r^2E^2+2f(r)(L^2+r^2)]rf'(r)+4f^2(r)L^2\right]f'(r)\\\nonumber
&&+2[-r^2E^2+f(r)(L^2+r^2)]f''(r)r^2f(r)+2f'^2(r)r^2f(r)(r^2+L^2)\\\nonumber
&&-4L^2f^2(r)(-f(r)\frac{3}{2}+r f'(r))].
\end{eqnarray}

\section{Accretion Regime}~\label{sec4}

Herein, we shall formulate some general results of accretion. For these results, one can use the energy-momentum tensor of a perfect fluid
\begin{equation}
T^{ab} =(\rho+p)u^a u^b + pg^ {a b},\label{32}
\end{equation}
with the following four-velocity vectors
\begin{equation}
u^a =(u^t,u^r,0,0)= \frac{dx^a}{d\tau},\label{33}
\end{equation}
where $\tau$ represents the particles geodesic motion proper time. For the analysis of normalization criterion for spherically symmetric steady-state flow is $ u^a u_a = -1$. We consider the four velocities of the present fluid as:
\begin{equation}
u^t =\frac{(f(r)+(u^r)^2)^{1/2}}{f(r)}.\label{34}
\end{equation}

The inward flow $u^r<0$ generates the negative velocity of the fluid and accretion. Whereas for the outward flow $u^r>0$, the fluid velocity should be positive. Therefore, for the present accretion analysis, we rely on the conservation law of energy and momentum tensor, which is further defined as:
\begin{equation}
T^{ab}_{;a} =0\Rightarrow T^{ab}_{;a}=\frac{1}{\sqrt{-g}}(\sqrt{-g}T^{ab})_{,a}+T^{\alpha a}\Gamma^{b}_{\alpha a}=0.\label{35}
\end{equation}

Now, by using the second kind of Christoffel symbol $\Gamma$ and the covariant derivative $;$, with $\sqrt{-g}=r^2\sin^2 \theta$, one can get the following relation
\begin{equation}
r^2u^r(\rho+p)\sqrt{(u^r)^2+f(r)}=D_0,\label{36}
\end{equation}
where $D_0$ is the  constant of integration. Now, by taking the befits from the relation of four-velocity and conservation law and through $u_{a}T^{ab}_{;b}=0$, we have
\begin{equation}
(\rho+p)_{;b}u_{a}u^{a}u^{b}+(\rho+p)u^{a}_{;b}u_{a}u^{b}+(\rho+p)u_{a}u^{a}u^{b}_{;b}+p_{,b}g^{ab}u_{a}+p u_{a}g^{ab}_{;b}=0.\label{37}
\end{equation}
By the supposition $ u^a u_a = -1$ and $g^{ab}_{;b}=0$, the above relation reduces to
\begin{equation}
(\rho+p)u^{b}_{;b}+u^{b}\rho_{b}=0.\label{38}
\end{equation}

Taking only non-zero components, we obtain
\begin{equation}
\frac{\rho'}{\rho+p}+\frac{u'}{u}+\frac{2r}{r^2}=0.\label{39}
\end{equation}

By integration
\begin{equation}
u^{r} r^2\exp\int\frac{d\rho}{\rho+p}=-D_{1},\label{40}
\end{equation}
where $D_{1}$ is the new constant of integration. 

Further, the minus sign is taken on the right-hand side that is $u^r<0$, we get
\begin{equation}
(\rho+p)\sqrt{\left[f(r)+(u^r)^2\right]}\exp\left(-\int\frac{d\rho}{\rho+p}\right)=D_{2},\label{41}
\end{equation}
where $D_{2}$ is the again constant of integration. The above setup produces the mass flux, which is further defined as:
\begin{equation}
(\rho u^a)_{;a}\equiv\frac{1}{\sqrt{-g}}(\sqrt{-g}\rho u^a)_{,a}=0,\label{42}
\end{equation}
and it is also revised as
\begin{equation}
\frac{1}{\sqrt{-g}}(\sqrt{-g}\rho u^a)_{,r}+\frac{1}{\sqrt{-g}}(\sqrt{-g}\rho u^a)_{,\theta}=0.\label{43}
\end{equation}

Therefore, we have the following conservation law for mass as
\begin{equation}
r^2\rho u^r=D_{3},\label{44}
\end{equation}
where $D_{3}$ is the constant like $D_{1}$ and $D_{2}$.

\subsection{Dynamical parameters}~\label{Dynamic}

In this Subsection, we shall suppose equation of state $p=k\rho$, where $k$ is the state parameter for the isothermal fluid. In is necessary to mention that in an isothermal fluids, the temperature will be constant throughout the flow. The sound speed parameters should be remained constant for such kind of fluids $p\propto\rho$ throughout the accretion. Now, from Eqs. (\ref{40}), (\ref{41}) and (\ref{44}), we have
\begin{equation}
\left(\frac{\rho+p}{\rho}\right)\sqrt{\left[(u^r)^2+f(r)\right]}\exp\left(-\int\frac{d\rho}{\rho+p}\right)=D_{4}.\label{45}
\end{equation}

In the above equation $D_{4}$ represents the integration constant. Now, by taking $p=k\rho$ in above equation, one can obtain the general form of radial velocity as
\begin{equation}
u(r)=\left(\frac{1}{k+1}\right)\sqrt{\frac{(D_{4})^2}{f(r)}-(k+1)^2}.\label{46}
\end{equation}

Therefore, the  final form of radial velocity for current considered AdS BH is given as
\begin{eqnarray}\label{47}
u(r)&=&\left(\frac{1}{k+1}\right)\sqrt{\frac{(D_{4})^2}{\left(1-\frac{2M}{r}+\frac{q}{3r}{\rm ArcSinh}\frac{q}{\sqrt{B}r^3}-\frac{r^2}{3}\sqrt{B+\frac{q^2}{r^6}}-\frac{r^2}{3\Lambda}\right)}-(k+1)^2}.
\end{eqnarray}

By considering Eq. (\ref{44}), the general form of the density for the relevant fluid is defined as
\begin{equation}\label{49}
\rho(r)=\frac{D_{3}}{r^2}\frac{(k+1)}{\sqrt{{\frac{(D_{4})^2}{f(r)}-(k+1)^2}}}.
\end{equation}

Now, the energy density for strong and weak fields is provided as
\begin{eqnarray}\label{50}
\rho(r)&=&\frac{D_{3}}{r^2}\frac{(k+1)}{\sqrt{{\frac{(D_{4})^2}{\left(1-\frac{2M}{r}+\frac{q}{3r}{\rm ArcSinh}\frac{q}{\sqrt{B}r^3}-\frac{r^2}{3}\sqrt{B+\frac{q^2}{r^6}}-\frac{r^2}{3\Lambda}\right)}-(k+1)^2}}}.
\end{eqnarray}

\begin{figure}
\centering \epsfig{file=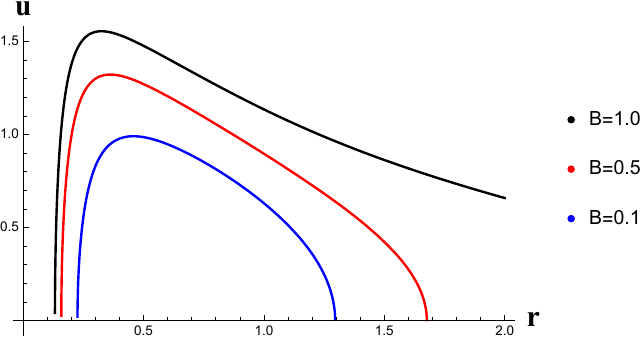, width=.45\linewidth,
height=2.05in}~~~~~~~~\epsfig{file=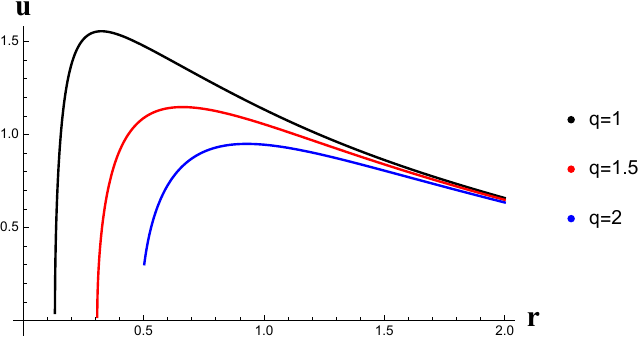, width=.45\linewidth,
height=2.02in}\caption{\label{F9} An illustration of the radial velocity of the AdS BH versus $r$ for different values of $B$ and $q$.}
\end{figure}

We characterized the velocity of fluid in Fig. \ref{F9} and has the following key points:
\begin{itemize}
  \item The fluid velocity is increased for increasing $B$. The bound radius is decreased when the velocity increases.
  
  \item The fluid velocity is increased for increasing $q$. Similarly, the bound radius is decreased when the velocity increases.
\end{itemize}

\begin{figure}
\centering \epsfig{file=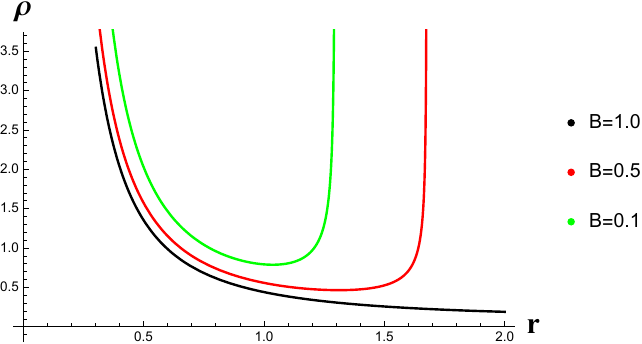, width=.45\linewidth,
height=2.05in}~~~~~~~~\epsfig{file=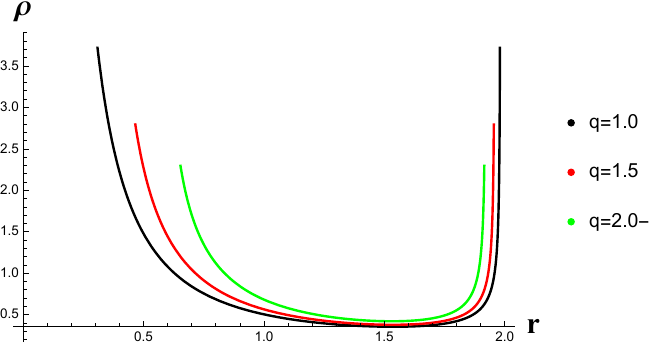, width=.45\linewidth,
height=2.02in}\caption{\label{F10} An illustration of the energy density of the AdS BH versus $r$ for different values of $B$ and $q$.}
\end{figure}

We characterized the density behavior in Fig. \ref{F10} and has the following key points:
\begin{itemize}
  \item The fluid density is decreased for increasing $B$, until it goes up near to the singularity $r=2.0$ at $B=1.0$.
  
  \item The fluid density is increased for decreasing $q$, and all the solution curves quickly fall onto the singularity.
\end{itemize}

\begin{figure}
\centering \epsfig{file=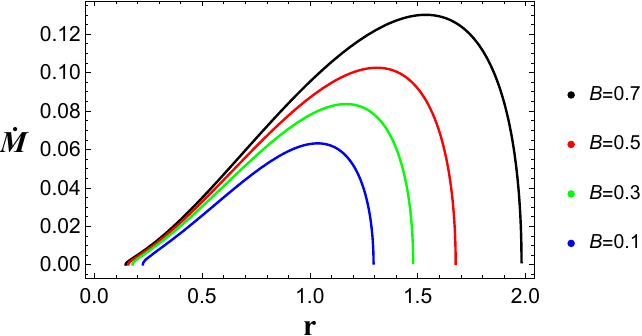, width=.45\linewidth,
height=2.05in}~~~~~~~~\epsfig{file=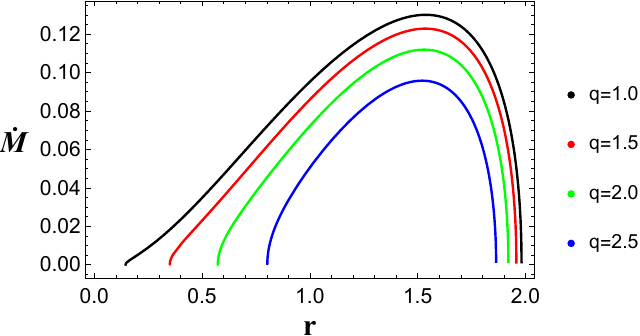, width=.45\linewidth,
height=2.02in}\caption{\label{F11} An illustration of mass accretion rate of the AdS BH versus $r$ for different values of $B$ and $q$.}
\end{figure}

We analyzed the mass accretion rate for AdS BH in Fig. \ref{F11} and has the following key points:
\begin{itemize}
  \item The maximum accretion rate of AdS BH occurs for ($B=0.7$) and ($q=1.0$). The radius is decreased gradually as increasing the accretion rate. 
  
  \item Accretion rate is increased for increasing $B$ and decreasing $q$ at different distances.
  \end{itemize}

\subsection{Mass expansion}~\label{Mass}

The mass of BH per unit of time can be measured by the mass accretion rate, which is equal to the area times flux at the boundary of BH. Further, it is usually denoted by $\dot{M}$, and it can be determined by the metric parameters as well as the characteristics of the accreting fluid. It is defined as:
\begin{equation}
\dot{M}=-4\pi r^2 u^r(\rho+p)\sqrt{(u^r)^2+f(r)}\equiv-4\pi D_0,\label{52}
\end{equation}
where $D_0=-D_1N_2$ and $D_2=\sqrt{f(r_\infty)}(\rho\infty+p_\infty)$ gives
\begin{equation}
\dot{M}=4\pi D_{1}\sqrt{f(r_\infty)}(\rho\infty+p_\infty)M^2.\label{53}
\end{equation}

In order to investigate the mass expansion, we should supposed that our boundary at infinity, i.e., $r=r_{\infty}$. The peculiar infinity where the massive particles fall into the BH is indicated by the radius at infinity. Also, $\rho_{\infty}$ is $\rho$ energy density at $r_{\infty}$ and $p_{\infty}$ means $p$ pressure at $r_{\infty}$. Hence, the above relation can be revised as:
\begin{equation}
\frac{dM}{M^2}=  \mathcal{F}dt,\label{54}
\end{equation}
where $\mathcal{F}=4\pi N_{1}(\rho_{\infty}+p{_\infty})\sqrt{f(r_{\infty})}$. 

On integrating, on can get
\begin{equation}
M_t=\frac{M_i}{1-Ft M_i}\equiv\frac{M_i}{1-\frac{t}{t_{cr}}},\label{55}
\end{equation}
where $M_i$ is the initial mass, $M_t$ is the mass of BH with the critical accretion time and $t_{cr}=\left[4\pi D_{1}(\rho_{\infty}+p{_\infty})\sqrt{f(r_{\infty})M_i}\right]^{-1}$ is the critical accretion of time evolution. Therefore, the final expression for BH mass accretion rate is expressed as:
\begin{equation}
\dot{M}=4\pi D_{1}(\rho+p)M^2.\label{56}
\end{equation}

\begin{figure}
\centering \epsfig{file=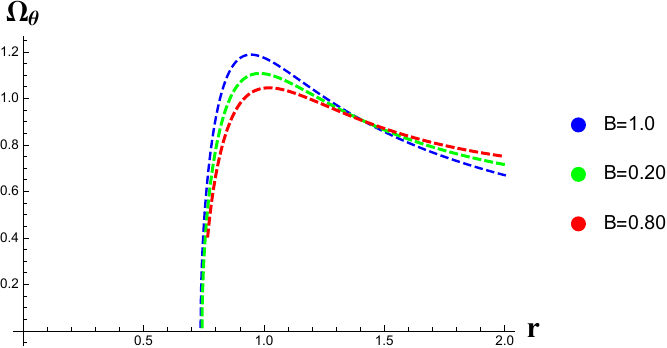, width=.47\linewidth,
height=1.9in}~~~~~~~~\epsfig{file=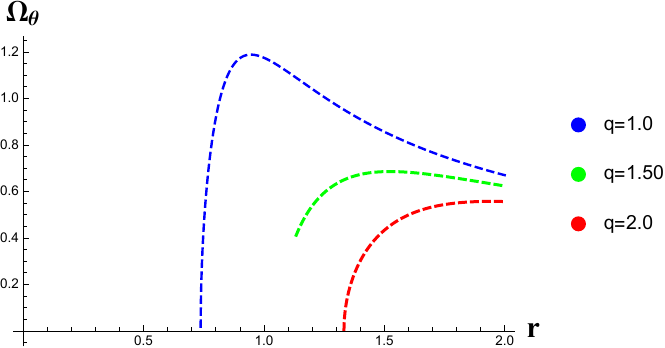, width=.47\linewidth,
height=1.9in}\\ \hspace{4cm}
\centering \epsfig{file=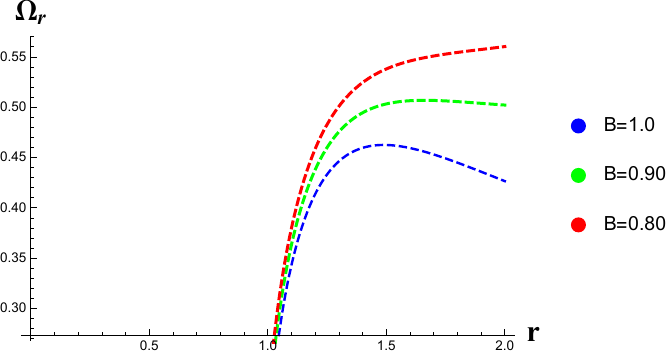, width=.47\linewidth,
height=1.9in}~~~~~~~~\epsfig{file=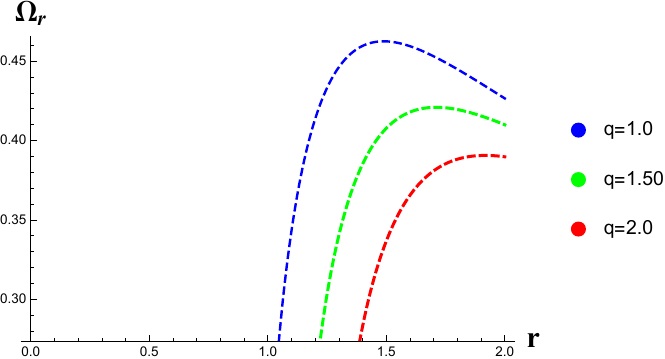, width=.47\linewidth,
height=1.9in}\\ \hspace{4cm}
 \epsfig{file=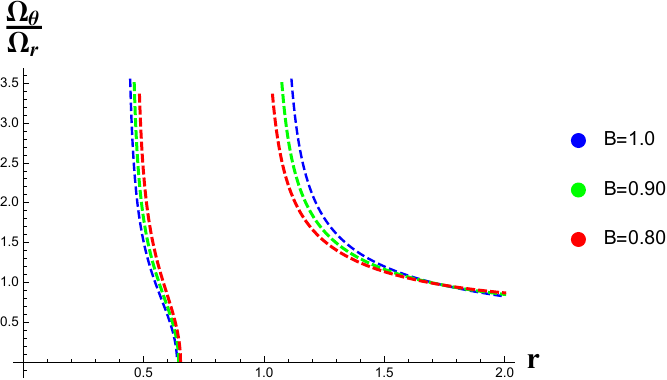, width=.47\linewidth,
height=1.9in}~~~~~~~~\epsfig{file=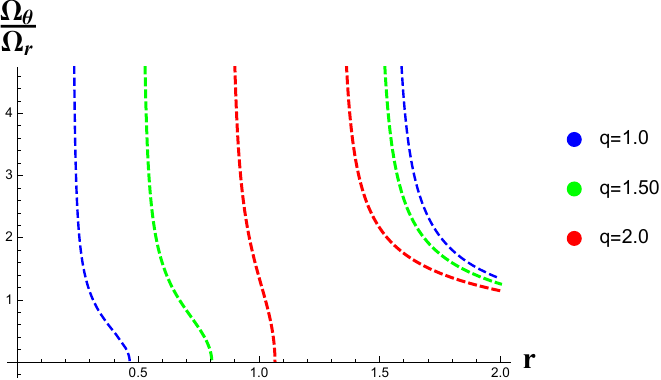, width=.47\linewidth,
height=1.9in}
\caption{\label{F12} An illustration of epicyclic frequencies of the AdS BH versus $r$ for different values of $B$ and $q$.}
\end{figure}

\section{Circular equatorial geodesics}~\label{sec5}

The explicit form of the effective potential, specific energy, specific angular momentum of a moving particle is important for the physical analysis.
From Eqs. (\ref{13}), (\ref{14}) and (\ref{15}), we have
\begin{eqnarray}\label{58}
V_{eff}&=&\left(1-\frac{2M}{r}+\frac{q}{3r}{\rm ArcSinh}\frac{q}{\sqrt{B}r^3}-\frac{r^2}{3}\sqrt{B+\frac{q^2}{r^6}}-\frac{r^2}{3\Lambda}\right)\left(1+\frac{L^2}{r^2}\right).
\end{eqnarray}
and
\begin{eqnarray}\label{64}
E^2&=&\frac{2\left(1-\frac{2M}{r}+\frac{q}{3r}{\rm ArcSinh}\frac{q}{\sqrt{B}r^3}-\frac{r^2}{3}\sqrt{B+\frac{q^2}{r^6}}-\frac{r^2}{3\Lambda}\right)^2}
{-\frac{6M}{r}+\frac{2Br^6+q^2\left[2+\left(-\sqrt{B+\frac{q^2}{r^6}}+\sqrt{B}\sqrt{1+\frac{q^2}{Br^6}}\right)r^2\right]}{q^2+Br^6}+\frac{q ~\text{arcsinh}\Big(\frac{q}{\sqrt{B}r^3}\Big)}{r}}.
\end{eqnarray}

\begin{eqnarray}\label{66}
L^2&=&\frac{2M+\frac{q^2}{\sqrt{B+\frac{q^2}{r^6}}r^3}-\frac{q^2}{\sqrt{B}\sqrt{B+\frac{q^2}{r^6}}r^3}-\frac{2\Lambda r^3}{3}-\frac{2}{3}\sqrt{B+\frac{q^2}{r^6}}r^3-\frac{1}{3}q ~\text{arcsinh}\Big(\frac{q}{\sqrt{B}r^3}\Big)}{-\frac{6M}{r}+\frac{2Br^6+q^2\left[2+\left(-\sqrt{B+\frac{q^2}{r^6}}+\sqrt{B}\sqrt{1+\frac{q^2}{Br^6}}\right)r^2\right]}{q^2+Br^6}+\frac{q ~\text{arcsinh}\Big(\frac{q}{\sqrt{B}r^3}\Big)}{r}}.
\end{eqnarray}

\subsection{Epicyclic frequencies}~\label{Epicyclic}

The moving particles in a circular orbit achieve a small oscillations in the direction of vertical and radial frequencies. These oscillations are the special effects of perturbation on the moving particles in a circular orbit. Hence, the required frequencies are given by
\begin{eqnarray}\label{72}
\Omega^2_{\theta}&=&\frac{-\frac{q^2}{\sqrt{B+\frac{q^2}{r^6}}r^3}-\frac{3q^2}{\sqrt{B}\sqrt{B+\frac{q^2}{r^6}}r^3}-6Mr^3+2\Lambda r^6-\sqrt{2Br^6+\frac{q^2}{r^6}}r^3+qr^3 ~\text{arcsinh}\Big(\frac{q}{\sqrt{B}r^3}\Big)}{{6M-3r+\Lambda r^3+\sqrt{1+\frac{q^2}{Br^6}}r^2}{{+Br^6}-q ~\text{arcsinh}\Big(\frac{q}{\sqrt{B}r^3}\Big)}}.
\end{eqnarray}

\begin{eqnarray}\label{73}
\Omega^2_{r}&=&\frac{\frac{2q^2}{\sqrt{2B+\frac{q^2}{r^6}}r^3}-\frac{3q^2}{\sqrt{B+\frac{q^2}{r^6}}r^3}+2Mr^2+3\Lambda r^3-\sqrt{Br^6+\frac{q^2}{r^6}}r^3-2qr^3 ~\text{arcsinh}\Big(\frac{q}{\sqrt{B}r^3}\Big)}{2r^3\left({2M-\Lambda r^2+\sqrt{1+\frac{q^2}{Br^6}}r^2}{{+Br^3}+q ~\text{arcsinh}\Big(\frac{q}{\sqrt{B}r^3}\Big)}\right)}.
\end{eqnarray}


We characterized the radial and vertical frequencies and their ratios in Fig. \ref{F12} with the following key points:

\begin{itemize}
  \item The vertical frequency $\Omega_{\theta}$ increased by increasing $B$ among very small change in the radius. 
  
  \item The vertical frequency $\Omega_{\theta}$ increased by decreasing $q$ and decreasing the radius. 
  
  \item The radial frequency $\Omega_{r}$ increased by decreasing $B$ among very small change in the radius. 
  
  \item The radial frequency $\Omega_{r}$ increased by decreasing $q$ and decreasing the radius.
  
  \item The ratio $\frac{\Omega_{\theta}}{\Omega_{r}}$ transformed the solution curves from minimum radius to maximum radius.
\end{itemize}

\section{Conclusion}~\label{sec6}

The geodesic motion and accretion process of a test particle near an AdS BH surrounded by a dark fluid with a Chaplygin-like equation of state have investigated in the current analysis. Under this specified theme, we examined the equatorial plane and circular geodesics along with their stabilities, radiations energy flux, oscillations and orbits. The general form of the fluid accretion onto the AdS BH through dynamical analysis and mass expansion also has discussed. Some other interesting features like the effective potential, angular momentum, specific energy, radiation energy and epicyclic frequencies have been investigated. All the important highlights of the current analysis are listed as follows:

\begin{itemize}
  \item The effect of parameters on the horizon via metric function $f(r)$ has provided in Fig. \ref{F2}. We have observed that AdS BH has intriguing behavior; with larger values of all the parameters $M$, $B$ and $q$, all of the solution curves are moved inward toward an event horizon.
  
  \item The effective potential for the considered AdS BH has presented in Fig. \ref{F3}. As effective potential is very important for the geodesic motion to find the location of the circular orbits by the local extremum. The stability of the orbits is also depended on the lapse function and the angular momentum $L$. The point of extremum around the AdS BH at a distance of $r\approx2$ represents the orbits locations, which we have investigated. Here, an increase in potential corresponds to an increase in angular momentum $L$ and parameter $q$, whereas a reduction in parameter $B$ results in a decrease in potential.
  
  \item The effect of involved parameters on specific energy $E$ and angular momentum $L$ have shown in Figs. \ref{F4} and \ref{F5} respectively. 
  
  \item The graphical behavior of integrated factor $Z$ from radiation parameter, radiation energy $K(r)$ and temperature $T$ with Stefan’s constant of the considered AdS BH has given in Figs. \ref{F6} and \ref{F7} respectively. Further,  accreting efficiency parameter has provided in Fig. \ref{F8}. It has noted that the process releases gravitational energy in the falling elements. The radiating energy corresponding to the specific energy $E$, angular momentum $L$, and angular velocity is what causes the radiation energy flux over the accretion disk.
  
  \item The behavior of radial velocity, energy density, and mass accretion rate of the proposed, which has surrounded by dark fluid with Chaplygin-like equation of state have given in Figs. \ref{F9}, \ref{F10} and \ref{F11} respectively. The radial velocity has increased for the increasing values of $B$. Interestingly, the bound radius has decreased when the velocity increased. The fluid velocity has increased for increasing in $q$. Similarly, the bound radius has decreased when the velocity increased. It has observed that the maximum accretion rate of AdS BH occurs for ($B=0.7$) and ($q=1.0$). The radius has decreased gradually as increased the accretion rate. 
  
  \item The effect of parameters $B$ and $q$ through different values on epicyclic frequencies of the AdS BH versus $r$ has shown graphically in Fig. \ref{F12}. 
\end{itemize}

In conclusion, the current analysis of accretion disks around AdS black holes surrounded by dark fluid through a Chaplygin-like EOS has offered a significant and insightful observations regarding the behavior of these type of systems. It can be noticed that the stability, structure and energy transport mechanisms of the accretion disks are modified by the emergence of new dynamics and attributes brought about by the existence of a dark fluid with an EOS  exhibiting the Chaplygin gas.

\section*{Data Availability}             
No new data were generated in support of this research. 

\section*{Conflict of Interest} 
The authors declare no conflict of interest.

\section*{Acknowledgement}
 G. Mustafa is very thankful to Prof. Gao Xianlong from the Department of Physics, Zhejiang Normal University, for his kind support and help during this research. Further, G. Mustafa acknowledges Grant No. ZC304022919 to support his Postdoctoral Fellowship at Zhejiang Normal University. The author SKM is thankful for continuous support and encouragement from the administration of University of Nizwa. SR gratefully acknowledges support from the Inter-University Centre for Astronomy and Astrophysics (IUCAA), Pune, India under its Visiting Research Associateship Programme and the facilities under ICARD, Pune at CCASS, GLA University, Mathura. This research is partly supported by Research Grant F-FA-2021-510 of the Uzbekistan Ministry for Innovative Development.\\

\end{document}